\documentclass[conference]{IEEEtran}
\IEEEoverridecommandlockouts

\usepackage{cite}
\usepackage{amsmath,amssymb,amsfonts}
\usepackage{algorithmic}
\usepackage{booktabs}
\usepackage{url}
\usepackage{graphicx}
\usepackage{textcomp}
\usepackage{xcolor}
\def\BibTeX{{\rm B\kern-.05em{\sc i\kern-.025em b}\kern-.08em
    T\kern-.1667em\lower.7ex\hbox{E}\kern-.125emX}}
\begin{document}

\title{AC-Mix: Self-Supervised Adaptation for Low-Resource Automatic Speech Recognition using Agnostic Contrastive Mixup \thanks{This work has been submitted to the IEEE for possible publication. Copyright may be transferred without notice, after which this version may no longer be accessible.}
}

\author{\IEEEauthorblockN{Carlos Carvalho and Alberto Abad} \IEEEauthorblockA{INESC-ID \& Instituto Superior T\'{e}cnico, Lisbon, Portugal\\
Email: carlos.mf.carvalho@inesc-id.pt and alberto.abad@inesc-id.pt}}

\maketitle

\begin{abstract}
Self-supervised learning (SSL) leverages large amounts of unlabelled data to learn rich speech representations, fostering improvements in automatic speech recognition (ASR), even when only a small amount of labelled data is available for fine-tuning. Despite the advances in SSL, a significant challenge remains when the data used for pre-training (source domain) mismatches the fine-tuning data (target domain). To tackle this domain mismatch challenge, we propose a new domain adaptation method for low-resource ASR focused on contrastive mixup for joint-embedding architectures named AC-Mix (agnostic contrastive mixup). In this approach, the SSL model is adapted through additional pre-training using mixed data views created by interpolating samples from the source and the target domains. Our proposed adaptation method consistently outperforms the baseline system, using approximately 11 hours of adaptation data and requiring only 1 hour of adaptation time on a single GPU with WavLM-Large.
\end{abstract}


\begin{figure*}[ht!]
  \centering
  \includegraphics[width=0.55\textwidth]{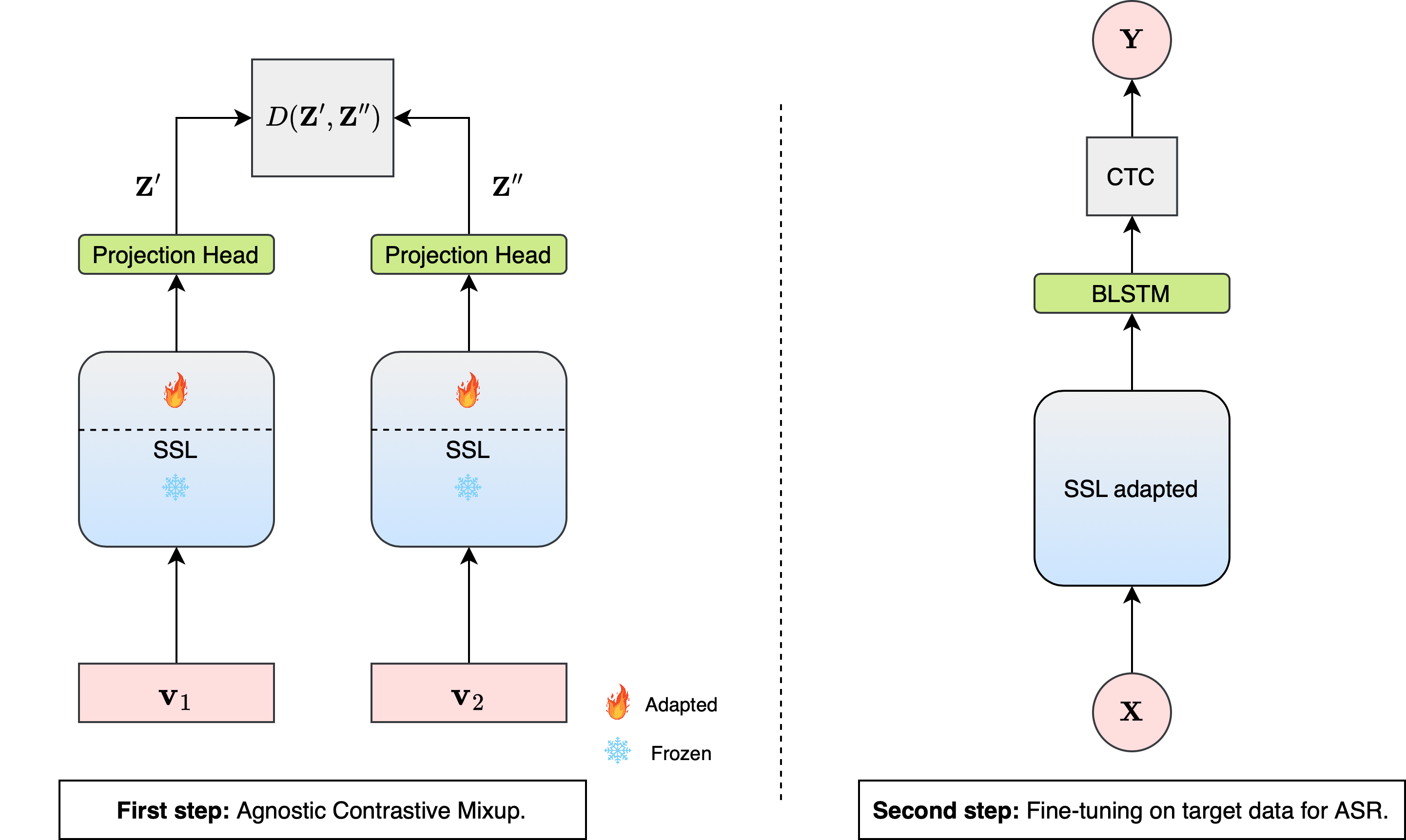}
  \caption{Two main steps of the AC-Mix method.}
  \label{fig:main_model}
\end{figure*}

\section{Introduction}

Recent advances in end-to-end (E2E) automatic speech recognition (ASR) stem largely from new deep learning architectures, like Transformer-based models \cite{vaswani2023attention, gulati2020conformer, rekesh2023fast}. However, these models require substantial labelled data and computational resources to perform well, posing challenges in low-resource settings \cite{radford2022robust, chen2021gigaspeech, chan2021speechstew}. In contrast, self-supervised learning (SSL) models like WavLM \cite{Chen_2022} leverage vast amounts of unlabelled data, which is easier to collect, and can capture rich representations that improve ASR performance and convergence speed, especially for low-resource settings \cite{Mohamed_2022}.

Despite advancements in SSL speech models, domain mismatch remains a major challenge. This domain bias occurs when fine-tuning data (target data) differs significantly from pre-training data (source data). Although using target domain data during pre-training improves ASR performance \cite{hsu2021robust}, this is feasible only for large companies or labs with extensive computational resources, enabling them to re-train large SSL models from scratch. Furthermore, much of the progress in speech SSL focuses on clean, non-spontaneous corpora like LibriSpeech \cite{baevski2020wav2vec, baevski2022data2vec, ma2023mt4ssl}, which may not address challenges in more complex and low-resource 
datasets.\cite{sanabria2023edinburgh,pradhan2023sciencetutormyst}


Addressing the domain bias of SSL models is crucial for improving speech recognition performance. Various methods have been proposed for this purpose \cite{lai2021parp, prasad2023pada, fan2022draft}. Data augmentation has proven to be crucial for domain bias compensation of SSL models in ASR when adapting to new data \cite{zaiem2023automatic}. Furthermore, SSL approaches based on joint-embedding training critically depend on data augmentation techniques to generate data views \cite{ng2023dehubert, jiang2021speech}. For instance, the approach proposed in \cite{chang2023spin} proposes a joint-embedding SSL system inspired by the SwAV loss \cite{caron2021unsupervised} to create speaker-invariant representations. To achieve this invariance, one view is augmented using a speaker perturbation method \cite{pmlr-v162-qian22b}, with the other being unaltered. In general, the effectiveness of these joint-embedding approaches largely depends on the right selection of the perturbations applied to the speech signals. 
Moreover, choosing these augmentations can be challenging due to the diverse nature of speech and the need to balance between introducing meaningful variations and avoiding the introduction of noise. 

In this work, we introduce a novel agnostic contrastive mixup (AC-Mix) domain SSL adaptation technique tailored for low-resource ASR. 
In particular, simple linear interpolation of samples from the source and the target domain, referred to as \emph{mixup}, is applied as a mechanism to generate alternative data views. 
Our proposed AC-Mix method consists of two main steps  (see Fig. \ref{fig:main_model}). 
In step one, we conduct SSL adaptation using AC-Mix on the joint-embedding system using only unlabelled data. In this phase, we explore different mixup strategies involving the source domain data (clean read speech), and the target domain data. 
 In the second step, we perform supervised fine-tuning using labelled target domain data combining the SSL model with a downstream prediction head. We evaluate our approach on WavLM base and large configurations, and additionally, we conduct an ablation study where we limit the number of supervised hours for  downstream ASR fine-tuning. Additionally, we report results for two different low-resource domains  



\section{Joint-embedding methods}
\label{sec:background_0part}

Joint-embedding methods (JEMs) usually optimize an encoder to produce similar embeddings for different augmented views of the same audio sample, using augmentations like speed perturbation and pitch shift, among others \cite{ng2023dehubert, jiang2021speech}. This enhances the encoder robustness to specific distortions and ensures invariance to augmentation. Furthermore, JEMs fall under the category of Energy-Based Models (EBMs) \cite{LeCun2006ATO} and often employ a contrastive learning framework. In this framework, the encoder minimizes the distance between embeddings of similar views while maximizing it for dissimilar views, preventing trivial solutions like constant outputs.

Several loss functions have been proposed for JEMs in the image domain, and some have been successfully adapted to speech, specifically for ASR, such as Speech SimCLR \cite{jiang2021speech} and Spin \cite{chang2023spin}. Particularly, the Spin framework exhibits rapid convergence, a trait observed in both image and speech domains. Unlike Speech SimCLR, which directly compares feature vectors, Spin soft-maps two feature vectors from different views $\mathbf{z}^{\prime}$ and $\mathbf{z}^{\prime\prime}$, to a set of K learnable clusters by creating their respective codes or mappings, $\mathbf{q}^{\prime}$ and $\mathbf{q}^{\prime\prime}$. Also, the batch is equally partitioned across clusters, ensuring distinct codes for different samples and avoiding trivial solutions where all samples receive the same code. A swapped prediction problem is then solved by predicting the code of one view using the representation of the other.

\section{AC-Mix approach}
\label{sec:method}


Several mixup-based strategies have been previously proposed for supervised E2E ASR \cite{ meng2021mixspeech, Xie_2023}.
However, while contrastive mixup has been explored in tasks like keyword spotting \cite{ng2023contrastive} and speaker recognition \cite{Zhang_2022}, to the best of our knowledge it has not yet been applied to SSL adaptation in low-resource ASR. To this end, we introduce a novel domain-agnostic augmentation method named Agnostic Contrastive Mixup (AC-Mix) inspired by \cite{xu2020adversarial}. The term "agnostic" here implies that the augmentation technique can be applied without needing detailed knowledge of the specific generation characteristics of the target data we aim to adapt to. Thus, the only prerequisite for our AC-Mix strategy is to have access to samples from the target distribution, $D_{\text{target}}$, and samples from the source data, $D_{\text{source}}$, used to pre-train the SSL system. As illustrated in Fig. \ref{fig:main_model}, our approach consists of two key steps: SSL adaptation with AC-Mix and subsequent ASR fine-tuning.

\subsection{SSL adaptation}
\label{sec:method_1part}

The first step, depicted on the left side of Fig. \ref{fig:main_model}, corresponds to the adaptation of the SSL model using unlabelled data. In AC-Mix, we employ a joint-embedding architecture along with the Spin loss mentioned in Section \ref{sec:background_0part} due to its efficiency and rapid convergence. In our proposed approach, 
we only consider two views  $\mathbf{v}_n$ ($n = \{1, 2\}$). In particular, the two alternative views of a sample $\mathbf{x}$ are generated by mixing $\mathbf{x}$ with two different randomly chosen samples $\mathbf{\tilde{x}}$. In this case, we simply interpolate the two audio samples:
\begin{equation}
\mathbf{v}_n = \lambda \mathbf{x} + (1 - \lambda) \mathbf{\tilde{x}_n},
\label{eq1}
\end{equation}
where $\lambda$ is a coefficient sampled from a uniform distribution $\lambda \sim U(\alpha, 1.0)$, with $\alpha$ values chosen from the range $[0, 0.9]$ in $0.1$ increments. The two mixed views of each training sample go through the shared SSL encoder, and subsequently, the resulting embeddings from the SSL encoder pass through a projection head, producing $\mathbf{Z}^{\prime}$ and $\mathbf{Z}^{\prime\prime}$, respectively.
Next, $\mathbf{Z^{\prime}}$ and $\mathbf{Z^{\prime\prime}}$ go through the Spin loss $D({\mathbf{Z}^{\prime}, \mathbf{Z}^{\prime\prime}})$ detailed in \cite{chang2023spin}. Similar to SwAV, Spin is a contrastive loss \cite{caron2021unsupervised}, making our mixup strategy inherently contrastive.

We propose 4 different strategies for mixing up the samples from the source distribution  $D_{\text{source}}$ with data from the target distribution  $D_{\text{target}}$.  
For each strategy, there are differences in the data utilized for SSL adaptation (i.e., the composition of the training batches) and the method used to generate the two alternative views as described in Equation \ref{eq1}:

\begin{itemize}
\item[] \textbf{Mixup1} Training batches consist of both $D_{\text{source}}$ and $D_{\text{target}}$ samples; $\mathbf{\tilde{x}_1}$ and $\mathbf{\tilde{x}_2}$ represent random samples from the batch. 
\item[] \textbf{Mixup2} Training batches contain only $D_{\text{target}}$ samples; $\mathbf{\tilde{x}_1}$ and $\mathbf{\tilde{x}_2}$ are random samples from  $D_{\text{source}}$.   
\item[] \textbf{Mixup3} Training batches contain only $D_{\text{source}}$ samples; $\mathbf{\tilde{x}_1}$ and $\mathbf{\tilde{x}_2}$ are random samples from  $D_{\text{target}}$.   
\item[] \textbf{Mixup4} Like \textit{Mixup3}, but $\mathbf{\tilde{x}_1}$ and $\mathbf{\tilde{x}_2}$ are the same in both views ($\lambda$ differs).
 \end{itemize}

\noindent
Additionally, during SSL adaptation, we exclusively adapt the last layers of the SSL architecture to enhance content-based representations, which are typically more prominent in the upper layers of the SSL system \cite{chang2023spin}.

\subsection{Fine-tuning for ASR}
\label{sec:method_2part}

For the second step of our approach, we perform supervised ASR training using target data only. This training process employs the SSL as the upstream model and integrates a prediction head as the downstream model. Illustrated on the right side of Fig. \ref{fig:main_model}, the prediction head comprises a Bidirectional Long Short-Term Memory (BLSTM) architecture. The loss function employed in this step is the Connectionist Temporal Classification (CTC) loss.

In practice, we carry out two distinct fine-tuning strategies with specific objectives in mind. First, in the \textit{Head-FT} strategy, we fine-tune only the downstream prediction head to evaluate the effectiveness of various mixup strategies and compare them with other approaches. Once the optimal mixup strategy is identified, we implement the \textit{Full-FT} strategy, where both the prediction head and the adapted layers of the SSL model are fine-tuned for the best ASR performance. This approach results in a more efficient and effective SSL adaptation, particularly in low-resource scenarios.

\section{Experiments}

\subsection{Corpora}

Our experiments utilize LibriSpeech \cite{panayotov2015librispeech}  and EdAcc \cite{sanabria2023edinburgh} corpora. 
As source domain data, specifically utilized for SSL adaptation as outlined in Section \ref{sec:method}, we rely on the LibriSpeech train-clean 100 hours subset, which includes 28,539 utterances from 585 speakers.
For the target domain, data from the EdAcc corpus is used.
EdAcc features English dyadic conversations with diverse accents and replicates real-world acoustic conditions typical in ASR engine deployment \cite{sanabria2023edinburgh}. 
The original EdAcc development participation serves as the target domain training data both for SSL adaptation and  supervised ASR fine-tuning, while the EdAcc test set is solely used for reporting results. Additionally, we create a random subset from the target domain training data for validation. In terms of data statistics, the target training data comprises 9600 utterances with 60 speakers, the validation subset contains 935 utterances with 6 speakers, and the target test data includes 9289 utterances with 60 speakers. For all EdAcc transcriptions we remove all annotations enclosed in "$<$","$>$" (e.g. $<\text{breath}>$) and instances enclosed in parentheses "(())". Finally, to assess AC-Mix performance on other corpora we use the My Science Tutor (MyST) Children Speech Corpus \cite{pradhan2023sciencetutormyst}. For this we use the same training, validation and testing partitions as proposed in \cite{10447091}. For adaptation and supervised ASR fine-tuning, we utilize the full training set with around 100 hours.

\begin{table}[th]
\setlength{\tabcolsep}{3pt}
\caption{WERs [\%] for EdAcc target Test Data. Evaluation of WavLM-Base AC-Mixup Strategies}
  \label{tab:mixup_selection}
  \centering
 \begin{tabular}{cc|cc}
\toprule
Adaptation&  $\alpha$ &test set &+LM \\
\toprule
-&-&40.4&  35.4  \\
\hline
Mixup1 &0.8&40.3& 35.3  \\
\hline
Mixup2 &0.9&41.1&  35.8   \\
\hline
Mixup3 
&0.3&\textbf{39.5}&  \textbf{34.5}   \\
\hline
Mixup4 &0.2&39.6& 34.8   \\
\bottomrule
  \end{tabular}
\end{table}

\begin{table}[th]
\setlength{\tabcolsep}{4pt}
\caption{WERs [\%] for EdAcc target set set. Evaluation of AC-Mix Adaptation vs. Baselines}
  \label{tab:baselines}
  \centering
 \begin{tabular}{cc|cc}
\toprule
\bf SSL& \bf Adaptation &test set &+LM \\
\toprule
WavLM-Base &-&40.4&  35.4  \\
\hline
WavLM-Base &Spin&39.7&  34.8
\\
\hline
WavLM-Base &EdAcc+Spin&41.0&  35.6   \\
\hline
WavLM-Base &Libri+EdAcc+Spin&40.3& 35.1 
\\
\hline
WavLM-Base &AC-Mix&\textbf{39.5}&  \textbf{34.5}   \\
\toprule
WavLM-Large &-&28.3& 24.7  \\
\hline
WavLM-Large &Spin&28.2&  24.8   \\
\hline
WavLM-Large &EdAcc+Spin&29.5&  25.6   \\
\hline
WavLM-Large &Libri+EdAcc+Spin&29.4&25.7 
\\
\hline
WavLM-Large &AC-Mix&\textbf{27.5}&  \textbf{24.4}   \\
\bottomrule
  \end{tabular}
\end{table}

\begin{table*}[t]
\setlength{\tabcolsep}{4pt}
\caption{WERs [\%] of E2E ASR models on EdAcc target test data for extreme low-resource ssettings. For the AC-Mix results, the first superscript indicates significance compared to the baseline without adaptation and the second superscript with respect to the SPIN baseline. Symbol $\dagger$ denotes significance (p-value of 0.01), while $-$ denotes no significant differences.}
  \label{tab:low-resources}
  \centering
 \begin{tabular}{cc|cc|cc|cc|cc}
\toprule &\multicolumn{1}{c}{}
 &\multicolumn{2}{c}{Full Set}
&
\multicolumn{2}{c}{5 hours} &
\multicolumn{2}{c}{1 hour} &
\multicolumn{2}{c}{10 minutes}\\\cmidrule(l){3-4}\cmidrule(l){5-6}\cmidrule(l){7-8} \cmidrule(l){9-10}
 \bf SSL & \bf Adaptation &test set &+LM  &test set &+LM &test set &+LM &test set &+LM \\
\toprule
WavLM-Base&-& 36.1
   & 33.1 & \textbf{41.3}  & 38.3 & \textbf{50.2} & \textbf{45.2} & 81.3 & 71.4 \\
\hline
WavLM-Base&Spin& 36.2 & 33.3  & 41.9  & 38.4 & 51.5 & 45.8 & 82.0 & 71.9 \\
\hline
WavLM-Base&AC-Mix& $\textbf{35.8}^{\dagger,\dagger}$ & $\textbf{32.8}^{\dagger,\dagger}$ & $41.4^{-,\dagger}$ & $\textbf{38.0}^{-,\dagger}$ & $50.9^{-,-}$ & $45.5^{-,-}$ & $\textbf{79.0}^{\dagger,\dagger}$  & $\textbf{70.8}^{\dagger,\dagger}$ \\
\toprule
WavLM-Large&-&  24.6  & 23.1 &  27.7 & 24.8 & 39.2 & 31.9 & 60.2 & 48.9 \\
\hline
WavLM-Large&Spin& 24.2 & 22.7 & 26.8  & 24.6 & 37.1 & 30.5 & 56.4 & 45.4\\
\hline
WavLM-Large&AC-Mix&  $\textbf{23.7}^{\dagger,\dagger}$  & $\textbf{22.4}^{\dagger,\dagger}$ &  $\textbf{26.6}^{\dagger,-}$ & $\textbf{24.2}^{\dagger,-}$ & $\textbf{36.5}^{\dagger,\dagger}$ & $\textbf{30.3}^{\dagger,\dagger}$ & $\textbf{56.1}^{\dagger,\dagger}$ & $\textbf{45.3}^{\dagger,\dagger}$ \\
\bottomrule
  \end{tabular}
\end{table*}

\begin{table}[th]
\setlength{\tabcolsep}{2pt}
\caption{WERs [\%] for MyST Test Partition}
  \centering
 \begin{tabular}{cc|cc}
\toprule
SSL& Adaptation &test set &+LM \\
\toprule
-&-&13.5&  12.5  \\
\hline
WavLM-Large &Spin&12.9& 12.0  \\
\hline
WavLM-Large &AC-Mix&$\textbf{12.5}^{\dagger,\dagger}$& $\textbf{11.8}^{\dagger,\dagger}$  \\
\bottomrule
\end{tabular}
\end{table}

\subsection{Experimental Setup}

For the initial step of our method, we used the code released from Spin \footnote{\url{https://github.com/vectominist/spin}} provided by \cite{chang2023spin} and run experiments with the WavLM-Base plus and with the WavLM-Large models both trained with 94k hours \footnote{Checkpoints: \url{https://github.com/s3prl/s3prl}}. All of our experiments were conducted on a single NVIDIA RTX A6000 with 48GB of memory capacity. The WavLM-Base models were adapted using the same configuration as Spin. For the base model, only the last two layers of the Transformer were adapted, following Spin \cite{chang2023spin}, while for WavLM-Large, the last five layers were used. The learning rates for WavLM-Base and WavLM-Large were set to $10^{-4}$ and $10^{-5}$, respectively. For WavLM-Large, the learning rate was linearly increased from 0 to $10^{-5}$ over 2.5k steps and then linearly decayed to $10^{-7}$. Similarly, for WavLM-Base, the rate was increased from 0 to $10^{-4}$ over 2.5k steps before decaying to $10^{-6}$.

In the second step of our approach, we utilized the S3PRL toolkit \footnote{\url{https://github.com/s3prl/s3prl}}. Specifically, we adapted the ASR downstream recipe proposed for the Speech processing Universal PERformance Benchmark (SUPERB) \cite{yang21c_interspeech} to suit the EdAcc and MyST corpora. During supervised fine-tuning with the EdAcc target training data, utterances shorter than 2.5 seconds were excluded due to their negative impact on ASR performance, as noted in preliminary experiments. All EdAcc systems were trained for 150k steps using Adam optimizer, with learning rates of $3 \times 10^{-4}$ for base models and $8 \times 10^{-5}$ for larger models. For the MyST corpus, the system was trained for 44k steps with a learning rate of $1 \times 10^{-4}$. The batch size for EdAcc and MyST experiments is 32. Also, the learning rate search was conducted in the range of $10^{-3}$ to $10^{-6}$.

For experiments varying the amount of supervised fine-tuning data, training steps were reduced to 75k, 15k, and 2.5k steps, respectively. In all experiments, we performed inference using the KenLM \cite{heafield2011kenlm} toolkit with the official 4-gram language model (LM) from LibriSpeech. It is worth noting that for all experiments, the final model checkpoint is the one saved at the last step. Additionally, for all AC-Mix results, only the best alpha value (ranging from 0.0 to 0.9), as determined empirically using the validation set, is presented. 

\subsection{Mixup strategy assessment}

Table \ref{tab:mixup_selection} showcases the results obtained with the diverse mixup strategies mentioned in Section \ref{sec:method}, using \textit{Head-FT} with the WavLM-Base model. Upon comparing the proposed mixup strategies, we observed that Mixup3 outperformed the others, achieving a WER of 39.5\% without LM and 34.5\% with LM. As a result, we opted for Mixup3 as our designated mixup strategy for all subsequent experiments.

\subsection{Comparison with other baselines}

In Table \ref{tab:baselines}, we compare our top AC-Mix adaptation strategy with other baselines for both WavLM-Base and WavLM-Large models. Following the efficient approach used to select the best mixup strategy, we perform only \textit{Head-FT} fine-tuning. The first baseline involves using the original SSL pre-trained model without adaptation. The second baseline employs the speaker perturbation method proposed in Spin \cite{chang2023spin}.  
Using this method, we adapt the SSL models either by training either with only the target training data (\textit{EdAcc+Spin}), only the source data  (\textit{Spin}), or both combined (\textit{Libri+EdAcc+Spin}). 

Regarding the speaker perturbation approach, only the \textit{Spin} adaptation strategy improves WER for WavLM-Base and WavLM-Large compared to no adaptation, while \textit{EdAcc+Spin} and \textit{Libri+EdAcc+Spin} either worsen performance or maintain baseline levels. In contrast, our mixup technique provides the most significant improvement for both models, regardless of LM application. 

Notably our strategy requires only one parameter for interpolation: the mixup parameter. In contrast, Spin \cite{chang2023spin} requires a more complex processing strategy that depends on several speaker-dependent hyperparameters that may not be available. Furthermore, the adaptation process on Wavlm-Large is completed in approximately 1 hour, making it highly cost-efficient.

\subsection{Supervised ASR in extremely low resource scenarios}

In Table \ref{tab:low-resources}, we investigate the impact of our AC-Mix approach followed by a varying amount of supervised data used for the \textit{Full-FT} setting : "full-set" with around 11 hours, "5 hours", "1 hour", and "10 minutes".
These smaller supervised subsets are created from the target domain training data.  
Additionally, we perform a significance test using Matched Pairs Sentence-Segment Word Error test \footnote{\url{https://github.com/talhanai/wer-sigtest}} for AC-Mix. 

We observe that AC-Mix is consistently better compared to both no adaptation and the \textit{Spin} adaptation when using both base and large WavLM models. Notably, our AC-Mix mixup strategy consistently improves over \textit{Spin} adaptation in all cases. Interestingly, as the supervised training data decreases, our AC-Mix method demonstrates more significant improvements, especially for the larger WavLM model.  

At last, regarding the generability of the proposed approach to other target low-resource domains besides accented speech, Table \ref{tab:low-resources} reports results for the children MyST corpus.
Experimental results show that AC-Mix improves the WER performance  compared to no adaptation and \textit{Spin} adaptation when using the WavLM-Large architecture also in this domain. 

\section{Conclusions and Future Work}

In this study, we introduce a novel domain adaptation strategy for low-resource ASR, named AC-Mix. AC-Mix blends samples from SSL pretraining with those from the target dataset, generating two augmented views crucial for training a contrastive joint-embedding system. We showcase the relevance of AC-Mix in enhancing the performance of E2E ASR systems under low-resource conditions for two different conversational corpora with low training costs. Notably, we observe AC-Mix increasing effectiveness as the amount of supervised adaptation data decreases, especially for WavLM Large. Our future work includes investigating the AC-Mix approach using increased amounts of  unlabelled data for adaptation.

\bibliographystyle{IEEEtran}
\bibliography{mybib}

\end{document}